\documentclass[aps,jmp,reprint,prl,showpacs]{revtex4-1}
\usepackage{amsmath,color,graphicx,dcolumn,multirow,hhline,blindtext}

\newcommand{\gout}{\ensuremath{\Gamma_\textnormal{out}}}

\begin{document}
\preprint{ETH Zurich}

% Title of the article
\title{Heat dissipation and fluctuations in a driven quantum dot}

\author{A. Hofmann}
\email[]{andrea.hofmann@phys.ethz.ch}
\author{V. F. Maisi}
\author{J. Basset}
\author{C. Reichl}
\author{W. Wegscheider}
\author{T. Ihn}
\author{K. Ensslin}
\affiliation{Laboratory for Solid State Physics, ETH Zurich}
\author{C. Jarzynski}
\affiliation{Department of Chemistry and Biochemistry and Institute for Physical Science and Technology, University of Maryland}

\date{\today}

\begin{abstract}
While thermodynamics is a useful tool to describe the driving of large systems close to equilibrium, fluctuations
dominate the distribution of heat and work in small systems and far from equilibrium. We study the heat generated by
driving a small system and change the drive parameters to analyse the transition from a drive leaving the system close
to equilibrium to driving it far from equilibrium. Our system is a quantum dot in a GaAs/AlGaAs heterostructure
hosting a two-dimensional electron gas. The dot is tunnel-coupled to one part of the two-dimensional electron gas
acting as a heat and particle reservoir. We use standard rate equations to model the driven dot-reservoir system and
find excellent agreement with the experiment. Additionally, we quantify the fluctuations by experimentally test the
theoretical concept of the arrow of time, predicting our ability to distinguish whether a process goes in the forward or
backward drive direction.
\end{abstract}
 
\maketitle   % please do not remove
\section{Introduction}
%study entropy production \cite{koski_distribution_2013}
%statistics of dissipated energy in driven signle-electron transitions \cite{averin_statistics_2011}
While equilibrium thermodynamics
is powerful in describing the energy balance in large systems
\cite{reichl_modern_1980}, 
fluctuations need to be taken into account when
driving small systems away from equilibrium. Over the
past two decades, the understanding of non-equilibrium
thermodynamics has been enhanced by fluctuation theorems
such as the Jarzynski and Crooks relations 
\cite{jarzynski_nonequilibrium_1997,crooks_entropy_1999,crooks_nonequilibrium_1998},
and their quantum extensions 
\cite{campisi_fluctuation_2009,talkner_tasaki_2007,albash_fluctuation_2013,rastegin_jarzynski_2014}.
These results reveal that
fluctuations in a system driven arbitrarily far from equilibrium
can be related to a very basic equilibrium quantity,
the free energy of the system. Effectively they allow us to
write statements of the second law of thermodynamics, for
small systems, as equalities rather than inequalities. Experimental
tests have verified these predictions in the classical
\cite{collin_verification_2005,saira_test_2012,liphardt_equilibrium_2002,kung_irreversibility_2012,carberry_fluctuations_2004,blickle_thermodynamics_2006,wang_experimental_2002,douarche_experimental_2005,koski_distribution_2013,hofmann_equilibrium_2016} 
as well as the quantum regime 
\cite{an_experimental_2015,batalhao_experimental_2014}.
Although on average, the second law of thermodynamics is valid, it
might be violated in single realizations of a process in the
presence of fluctuations.

We study in detail the transition from equilibrium to
non-equilibrium in the example of dissipation caused by
single electrons in a semiconductor quantum dot coupled to
a reservoir. A suitable gate voltage can dynamically drive
the single-particle states of the quantum dot with respect to
the Fermi levels of the reservoirs. The dissipation is measured
for every single drive realization, which makes a statistical
analysis of the distribution possible and enables us
to characterize the fluctuations. Depending on whether the
drive is slow or fast compared to the tunnelling rates between
dot and reservoir, the process can be considered to be
equilibrium or non-equilibrium. We show how the resulting
distribution of dissipation resembles a Gaussian shape
near equilibrium, and we explain the less regular curve and
the origin of the sharp features away from equilibrium.
Finally, we use our experiments to verify a theoretical
prediction that, in effect, quantifies the ability to determine
the direction of the arrow of time, in small systems driven
away from equilibrium
\cite{jarzynski_equalities_2011}.

\section{Experimental setup}
\begin{figure}[htb]%
\includegraphics[width=\linewidth]{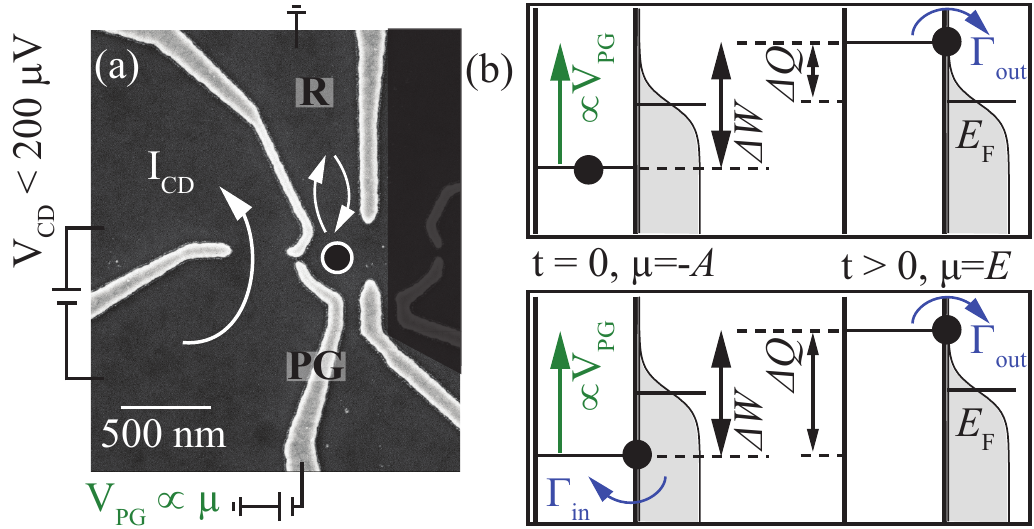}
\caption{Experimental setup. (a) shows a scanning electron micrograph of
a sample with the same lithographic design as was used for our experiment. 
An electron, drawn in black, can tunnel back and forth between the quantum dot
and the reservoir 'R'. The current $I_\textrm{CD}$ measures the occupation
of the dot. The schematics of the dot-reservoir system in (b) show how 
work and heat are defined in our experiment.
The upper panel shows a realization where the quantum
	dot is occupied in the beginning, hence $\Delta Q \leq A$. In the lower
	panel, the dot is empty at $t=0$, and a tunnelling event at small $t>0$
	time can lead to $|\Delta Q| \approx 2A$.
}
\label{fig:Setup}
\end{figure}

In our experiment, we grow a GaAs / AlGaAs heterostructure to
host a two-dimensional electron system 90 nm below the surface. 
We define a quantum dot by applying negative voltages to top-gates,
as shown in Fig.~\ref{fig:Setup}(a), thereby depleting the two-dimensional
electron gas below. The confinement gives rise to a spectrum with typical
single-particle level spacings of the order of $100~\mu$eV. 
The Coulomb repulsion between electrons in the quantum dot 
introduces an additional energy splitting
of the order of $1~$meV which prevents a second electron from entering 
the quantum dot at the same energy, and allows
for studying tunnelling processes of single electrons between the
dot and the reservoir. For this, we utilize the current $I_\textrm{CD}$ 
through a nearby channel, which is sensitive to changes in the occupation
of the quantum dot
\cite{schleser_time-resolved_2004,vandersypen_real-time_2004}.
The quantum dot is tunnel-coupled to 
one region of the two-dimensional electron system [labeled 'R' in Fig.~\ref{fig:Setup}(a)] 
acting as a heat and particle reservoir. The reservoir is described 
by a Fermi distribution with a Fermi energy $E_\textrm{F}$ and 
a temperature $T=40~$mK corresponding to an energy of $3.4~\mu$eV.
	The temperature has been extracted from the width of a Coulomb resonance,
	which measures the thermal broadening of the Fermi distribution in the reservoir
	\cite{beenakker_theory_1991,ihn_semiconductor_2009}.

The schematic illustrations in Fig.~\ref{fig:Setup}(b) show the lowest-lying
electrochemical potential $\mu$ of the quantum dot, 
together with the reservoir Fermi distribution. 
In thermodynamic equilibrium, if $\mu \ll E_\textrm{F}$, the state at 
$\mu$ is occupied, while it is empty for
$\mu \gg E_\textrm{F}$. If the energy of the dot level is within an energy 
window of about $4kT$ around the reservoir Fermi energy, the electron is allowed 
to statistically tunnel back and forth between the dot and the reservoir, at a tunnelling
rate which we tune to about $50$~Hz. Due to the two-fold spin-degeneracy of the lowest energy level
\cite{ciorga_addition_2000,tarucha_shell_1996}, 
the tunnelling-in rate, $\Gamma$, is twice as large as the tunnelling-out rate, 
$\gout = 2 \Gamma$
\cite{fujisawa_bidirectional_2006,maclean_energy-dependent_2007}.
By measuring the tunnelling rates in thermodynamic equilibrium we probe the Fermi function
	of the electron reservoir, which provides the energy scale $kT$
	in units of applied gate voltage, as described in \cite{hofmann_equilibrium_2016}.

\section{Heat and work in a driven quantum dot}
\begin{figure}[htb]%
\includegraphics[width=\linewidth]{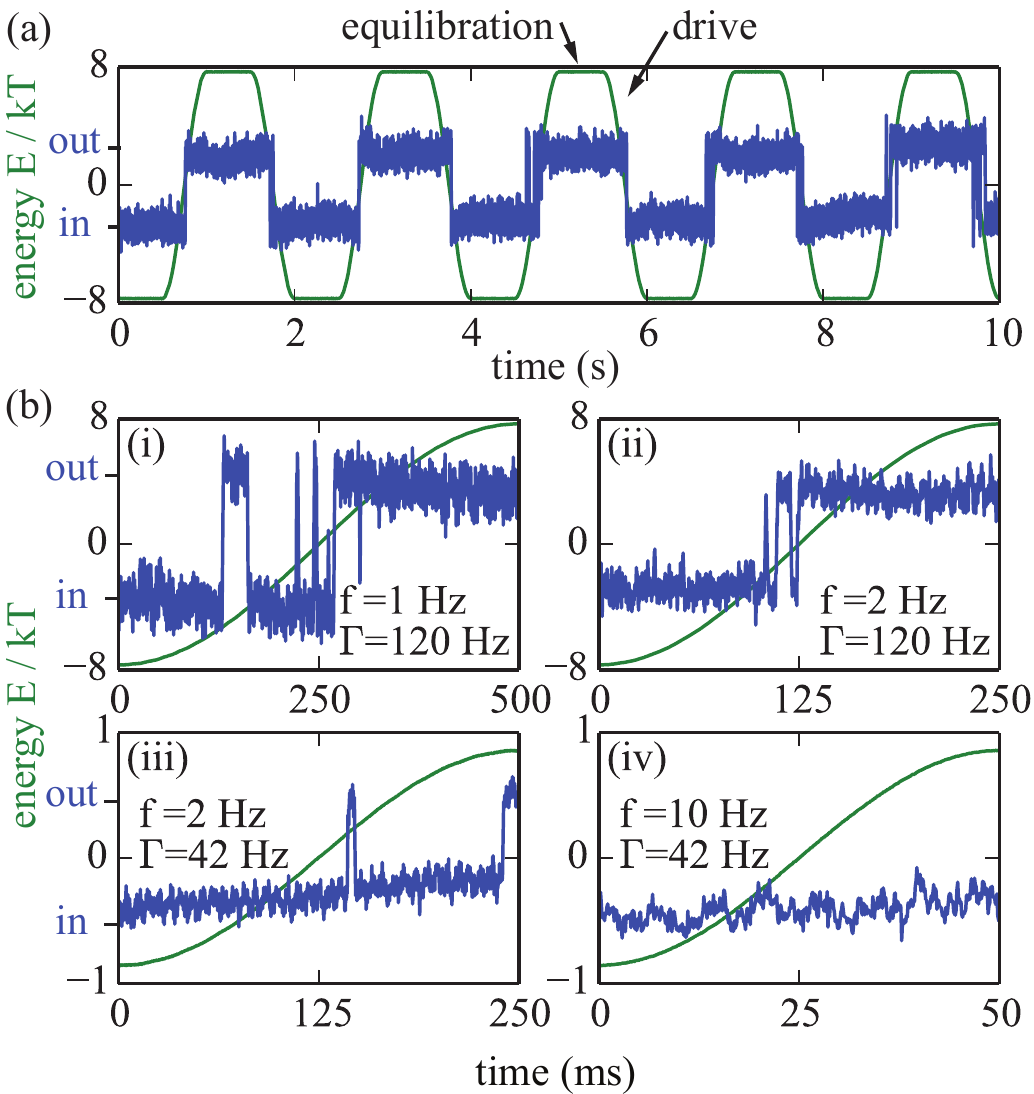}
\caption{Drive realizations. In (a), a full time trace of ten seconds shows
the energy of the quantum dot (green) being driven between $\pm8~kT$, as well as the 
equilibration periods in between. The charge detector current (blue) meanwhile
monitors the occupation of the dot. In (b), we plot realizations of the drive
for different frequencies, amplitudes and tunnelling rates.}
\label{fig:Traces}
\end{figure}

An arbitrary-wave form generator connected to the gate 'PG'[see Fig.~\ref{fig:Setup}(a)] 
shifts the energy of 
$\mu$ with a time-periodic wave-form shown in green in Fig.~\ref{fig:Traces}(a). 
We distinguish two sections in this wave-form, namely a waiting time of $0.5$~s 
(flat sections) which allows the quantum dot to reach thermal equilibrium 
with the reservoir, and the drive (sections with finite slope).
The drive consists of half a period of a sine with a frequency $f$ and an amplitude
$A$. Over the course of the entire measurement, we monitor the occupation 
(empty = 'out', occupied = 'in') of the dot with the charge detector current, as shown in blue in Fig.~\ref{fig:Traces}.

In the following, we analyse the dissipation and work attributed to driving the
quantum dot level from $\mu=-A$ to $\mu=A$. Many hundreds of time traces 
are recorded for a given parameter setting $f,A,\Gamma$. An example of such a 
time trace is shown in Fig.~\ref{fig:Traces}(a).
Small energy drifts of the quantum dot due to its environment are taken 
care of by readjusting gate voltages regularly such that the average dot occupation
stays constant. Different realizations of the dot occupation $q(E)$
for different parameter settings are plotted in Fig.~\ref{fig:Traces}(b). 

If the level is occupied, work $\Delta W$ is performed on the electron
in the dot
\cite{pekola_work_2012,saira_test_2012}
while the level is shifted in energy, as shown in Fig.~\ref{fig:Setup}(b).
Through an elastic tunnelling process, the electron can leave the dot and 
relax in the reservoir. The dissipation 
$\Delta Q$ in the relaxation process equals the energy difference
$E(t) - E_\textrm{F}$ between the dot level at the time $t$ of the tunnelling
event and the reservoir Fermi energy
\cite{pekola_work_2012,saira_test_2012}.
With the measured parameter $q(E)$ describing the occupation
(empty = 'out'  = 0, occupied = ' in' = 1) of the quantum dot 
at a certain energy $E$, as obtained from $I_\textrm{CD}$
traces, these considerations allow us to determine the total work
performed onthe quantum dot and the total dissipation in
the reservoir within one section of the drive by calculating
\begin{align}
\Delta W = \int_{E = -A}^{E = A} q(E) dE \\
\Delta Q = \int_{E = -A}^{E = A} \left( -\frac{dq}{dE} \right) E dE \label{eq:Heat}
\end{align}
The derivative $-dq/dE$ is non-zero only at energies, where a tunnelling event occurs,
and is positive for tunnelling-out and negative for tunnelling-in events.

In most cases, the dot is occupied when the drive starts at $\mu = -A$.
When $\mu$ is driven up in energy and comes closer to the reservoir Fermi energy, 
electron tunnelling events between the dot and the reservoir are observed, as is visible
in the charge detector signal (blue) in Fig.~\ref{fig:Traces}. The number of
events is larger for large tunnelling rate $\Gamma$ and slow drive frequency $f$.

\section{Heat distribution}
\begin{figure}[htb]%
\includegraphics[width=\linewidth]{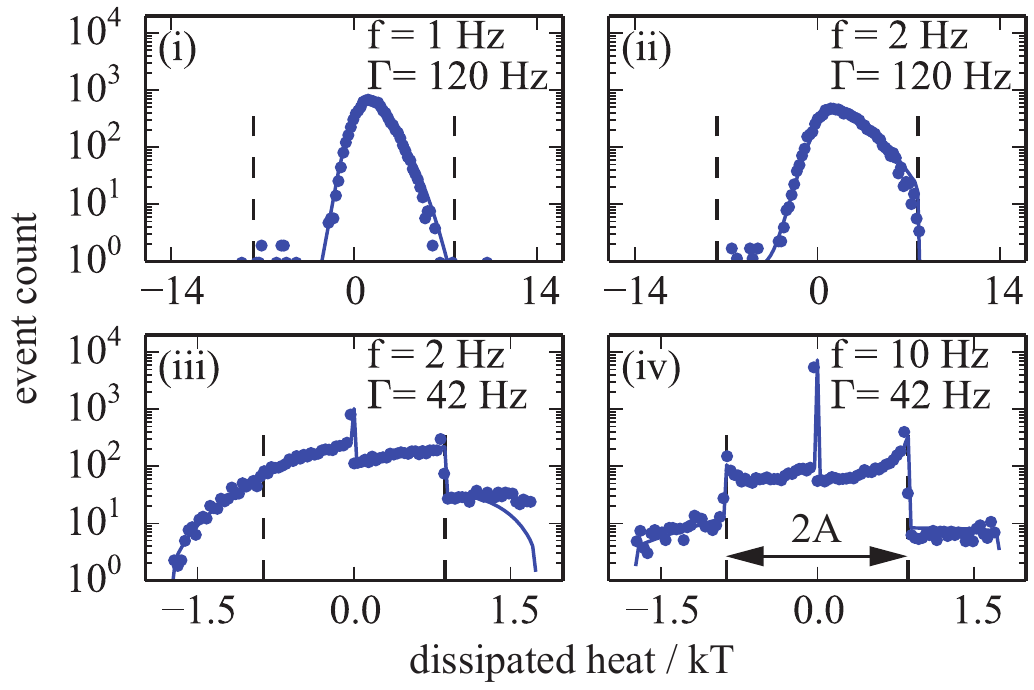}
\caption{Distributions of dissipated heat
for the upward (forward) drive direction and}	
drive settings as described in
Fig.~\ref{fig:Traces}(b).
The circles are experimental values, and the solid
lines are master-equation simulations without free parameters.
\label{fig:Distributions}
\end{figure}

The ensemble of several thousand realizations for each parameter setting allow for a 
statistical analysis of the dissipation 
\cite{koski_distribution_2013,averin_statistics_2011,hofmann_equilibrium_2016}. 
The individual realizations shown in 
Fig.~\ref{fig:Traces}(b) are typical traces corresponding to the distributions shown 
in the four panels of Fig.~\ref{fig:Distributions}. 

In Fig.~\ref{fig:Distributions}, the distribution shown in panel (i) has an almost 
Gaussian shape, as expected for a nearly static drive with large amplitude 
but low frequency.
After the equilibration period, at $\mu = -A$, the quantum dot is occupied with 
high probability. The low frequency drive and high tunnelling rate keeps the 
dot-reservoir system close to equilibrium during the full drive time,
leading to the Gaussian shape
\cite{koski_distribution_2013,averin_statistics_2011,douarche_experimental_2005}.
Due to the large number of tunnelling events, the dissipation scatters around
a mean value of zero, with the variance depending on the temperature
\cite{nyquist_thermal_1928}. 
With the dot occupied in the
beginning of the drive, the maximum possible dissipation is $\Delta Q = A$.

We compare the experimental distribution with a simple rate equation model
\cite{saira_test_2012}.
The model is fully defined by the drive parameters $f,A$, the tunnelling rate $\Gamma$
and the conversion factor between the voltage on 'PG' to energy. All these parameters
are known from the experiment, as described also in Ref.~\cite{hofmann_equilibrium_2016}.
The initial condition of the drive is set by the probability $p_\textrm{in}(E=-A)$ 
of the quantum dot level $\mu$ to be occupied at the energy $E=-A$, which can 
be directly calculated from the partition function $Z(E=-A)$ of the quantum dot.
The relevant single-particle energies are $\epsilon_0 = 0$ for 
the empty dot and $\epsilon_{1,2} = -A$ for each of the two spin-degenerate states 
where an electron occupies the dot. This gives
\begin{align}
\label{eq:partfun}
Z(E=-A) &= \sum_i e^{-\frac{\epsilon_i}{kT}} = 1 + 2 e^{\frac{A}{kT}} \\
p_\textrm{in}(E=-A) &= \frac{2}{Z} e^{\frac{A}{kT}} \\
p_\textrm{out}(E=-A) &= 1/Z
\end{align}
We plot the result from the rate equation simulation as solid lines in 
Fig.~\ref{fig:Distributions} and find excellent agreement to the measured data.

For the second distribution, shown in panel (ii), the drive amplitude A and tunnelling rate $\Gamma$ are equal to those in (i), but the drive frequency $f$ is doubled. This leads to an increased frequency of measuring positive dissipation: according to Eq.~\eqref{eq:Heat}, 
positive contributions to the dissipation result from electrons tunnelling 
into the quantum dot at negative energies ($\mu < E_\textrm{F}$), and 
electrons tunnelling out of the quantum dot at positive energies 
($\mu > E_\textrm{F}$). Considering that tunnelling-in (-out) 
only occurs if an occupied (free) state is available
in the reservoir, it becomes clear that positive dissipation is more probable.

For the distribution shown in panel (iii), the tunnelling
rate has been decreased by a factor of three, while the drive
frequency has been kept equal. Effectively, this decreases
the number of tunnelling events and drives the dot further
out of equilibrium with respect to the reservoir. As a result,
sharp features appear at $\Delta Q = 0,A$. 
While the peak at zero dissipation will be discussed below, let us 
concentrate first on the sharp step. A realization with
$\Delta Q \approx A$ is shown in Fig.~\ref{fig:Traces}(b)(iii). The
dot is occupied in the beginning and the electron tunnels out 
at high energy $E\approx A$, leading to 
dissipation of about $A$. 
The tunnelling events in-between are
close to each other and therefore do not contribute significantly
to the total dissipation. These realizations are likely
because of the fast drive, but also the shape of the drive:
the sinusoidal form enhances the probability for the electron
to tunnel out at high energy.
On the other hand, dissipation $\Delta Q > A$ 
occurs only in realizations where the dot
is empty in the beginning of the drive, which is suppressed by 
$p_\textrm{in} / p_\textrm{out} = 2 \exp(A/kT)$. Due to the lower drive amplitude in panel (iii)
compared to (i) and (ii), a significant number of events are observed with large dissipation.

Interestingly, the twofold degeneracy of the dot energy
level also influences the shape of the distribution.
The probabilities for an empty and occupied dot are equal
at $E=-kT \ln(2)$. Therefore, if the drive is symmetric
around the Fermi energy, the heat distribution has a different
shape for the drive upwards in energy than downwards
in energy, although all other drive parameters are equal, as
illustrated in 
\cite{hofmann_equilibrium_2016}.

The distribution shown in panel (iv) results from an
even faster drive of $f = 10$~Hz. In this case, realizations
with very few tunnelling events are likely. For example,
the peak at zero dissipation results from realizations like
the one shown in Fig.~\ref{fig:Traces}(b,iv), where no tunnelling event
occurs and $dq/dE= 0$ everywhere. 
On the other hand, realizations
with finite but small dissipation become unlikely,
because they require tunnelling events to occur at $\mu \sim E_\textrm{F}$,
where the drive is steepest. Hence, the peak in the distribution
is a delta-peak. Steps are visible at $\Delta Q = \pm A$, as
explained above, due to the sinusoidal waveform, which
favours tunnelling events at the beginning and end of the
drive; and due to the unlikely initial condition required for
$\Delta Q > A$. The distribution found here is very similar to
that found in Ref.~
\cite{hofmann_equilibrium_2016}, though it is measured with a different
drive amplitude.

\section{Arrow of time}
The distributions in Fig.~\ref{fig:Distributions} show the presence of fluctuations
in the heat dissipation in the dot-reservoir system which we study here. Using the
thermodynamic state variable $\Delta U$ describing the change in internal energy, we 
can obtain the work $\Delta W = \Delta U - \Delta Q$ performed during any realization of the process,
which allows us to study fluctuations in the work.
These fluctuations can be used to quantify the notion of the arrow of time,
as we describe in the following.
%With a large enough statistics, the Jarzynski equation allows us to calculate the change in
%free energy, $\Delta F$, between the initial and final state of the drive,
%\begin{align*}
%\Delta F = -kT \ln\left( \left\langle \exp\left(-\frac{\Delta W}{kT}\right) \right\rangle \right)
%\end{align*}
\begin{figure}[htb]%
\includegraphics[width=\linewidth]{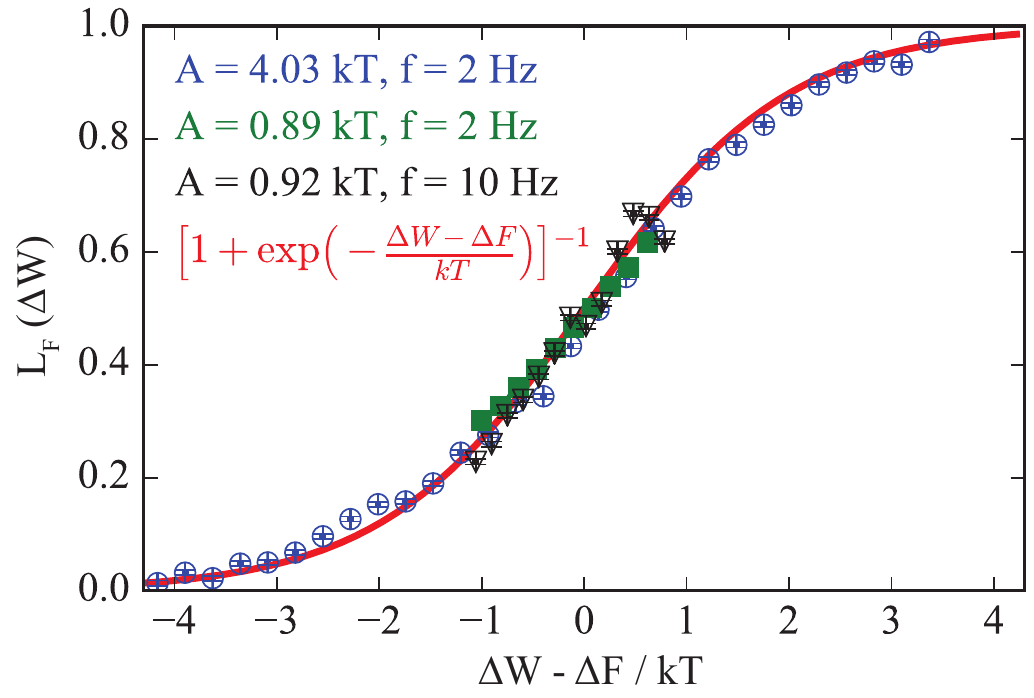}
\caption{Arrow of time: the plot shows the likelihood $L(F)$ that a 
specific value $\Delta W$ is measured in a realization in the 
forward direction. The green, blue and black markers show values
obtained from the experiment as described in Eq.~\eqref{eq:movieExp}
(with the error bars indicating the statistical error),
while the red line shows the theoretical prediction according to Eq.~\eqref{eq:movieTh}.
The green squares and black triangles correspond to the data shown in Fig.~\ref{fig:Distributions},
panels (iii) and (iv), respectively.
}
\label{fig:Movie}
\end{figure}

Imagine that we watch a movie of a system undergoing
an irreversible, isothermal process, and we must guess whether the movie is being run in the `forward' or the `backward' direction.
In this movie, we observe that an amount of work $\Delta W$ is performed on the system,
and the free energy between the initial and final states of the system is $\Delta F$.
For a macroscopic system, if the inequality $\Delta W > \Delta F$ is satisfied then we can state with certainty
that the movie is running forward in time, but if we observe $\Delta W < \Delta F$,
then we conclude with equal confidence that the movie is being run backward in time.
In other words, for large systems the arrow of time points in a direction specified by the second law of thermodynamics.

With small systems, where fluctuations dominate, this distinction becomes blurred.
Nevertheless, we can quantify our ability to determine the direction of time's arrow~\cite{jarzynski_equalities_2011}.
Specifically, given values of $\Delta W$ and $\Delta F$, the likelihood that the movie is being run forward is~\cite{shirts_equilibrium_2003,maragakis_bayesian_2008}:
\begin{align}
\label{eq:movieTh}
L(F) = \left[ 1+ e^{-\frac{(\Delta W-\Delta F)}{kT}} \right]^{-1}
\end{align}
When $|\Delta W - \Delta F| \gg kT$, this result gives $L(F)\approx 1$ or $L(F)\approx 0$, in agreement with the macroscopic case.
However, when $\Delta W$ is very close to $\Delta F$, $L(F)$ takes on an intermediate value, reflecting the difficulty
in distinguishing between forward and backward in time.
We have used our data to test Eq.~\ref{eq:movieTh}.

In our experiment, we drive the quantum dot in two directions, upwards ('forward') and downwards
('reverse') in energy, as shown in Fig.~\ref{fig:Traces}(a). 
Each cycle of driving thus contains one realization of forward driving, followed by one realization
of reverse driving.
For each realization in the reverse direction, we multiply the work by $-1$, to obtain the value that we would observe
if a movie of the process were run backward in time; in such a movie, it would appear that we are observing
a realization of the forward process.
By this procedure, from $N$ cycles we obtain $2N$ values of $\Delta W$,
all of which putatively represent realizations of the forward process.
Half of these work values are obtained from true realizations of the forward process; the other half are 'fakes', coming
from time-reversed realizations of the reverse process.

We binned both the forward (`true') and the backward (`fake') work values, letting $n_\textrm{fwd/bwd}$ denote
the number of forward/backward counts in a specified bin $\Delta W$.
For each bin, we then computed the fraction of forward counts:
\begin{align}
\label{eq:movieExp}
p_\textrm{fwd} (\Delta W) =  \frac{n_\textrm{fwd} (\Delta W)}{n_\textrm{fwd} (\Delta W) + n_\textrm{bwd} (\Delta W)}
\end{align}
This fraction represents the empirical likelihood $L(F)$ that a work value $\Delta W$
is due to the forward (rather than a backward) realization of the process under study.
The free energy difference has been computed analytically by utilizing the partition function 
of the system [Eq.~\eqref{eq:partfun}] and
the relation $F = - kT \log(Z)$.

In Fig.~\ref{fig:Movie}, we plot the theoretical prediction, Eq.~\eqref{eq:movieTh},
as a solid red line together with the probability $p_\textrm{fwd}$ obtained from experimental data
with different drive parameters, 
but the same tunnel coupling $\Gamma = 42$~Hz.
%In agreement with Eq.~\eqref{eq:movieTh}, the data collapse 
The data collapse onto a single curve given by Eq.~\eqref{eq:movieTh}, confirming
that the form of this curve depends only on the temperature, and not on other parameters such as
the amplitude and frequency of driving.
The error bar denotes the error due to the finite statistics of few thousand
	realizations. An additional systematic error occurs due to voltage drifts and the fluctuations
	in the electrostatic environment in the quantum dot.
	However, the good agreement between data and simulation
	shown in Fig.~\ref{fig:Distributions} suggest the
	systematic error to be small.

Mathematically, Eq.~\eqref{eq:movieTh} is essentially a consequence of Crooks's fluctuation theorem~\cite{crooks_entropy_1999,crooks_nonequilibrium_1998}.
Conceptually, however, it is rather remarkable that the ability to distinguish the direction of time's arrow
can be quantified by a formula as simple and universal as Eq.~\eqref{eq:movieTh}.

\section{Conclusion}
We utilize a GaAs/AlGaAs quantum dot for measuring heat distributions in a small system,
where fluctuations dominate and equilibrium predictions fail. By driving the quantum dot
with different drive parameters, we show how the distributions of heat dissipation can 
deviate largely from the Gaussian form that arises near equilibrium. We characterize the work
fluctuations in terms of the arrow of time and find that
in small systems, the fluctuations blur the distinction between forward and backward
realizations in a thermally broadened window of values for the work around the equilibrium free energy.
%\begin{equation}
%\label{eq1}
%\frac{a}{b}=\frac{c}{d}
%\end{equation}

%\section{Section}

%\subsection{Subsection}

%\subsubsection{Subsubsection}

%\paragraph{Paragraph}

%\subsection{Another subsection}

\begin{acknowledgements}
We thank Jukka Pekola and Ivan Khaymovich for stimulating discussions.
We greatfully acknowledge the U.S. National Science Foundation under grant DMR-1506969,
and QSIT via the SNF for providing the funding which enabled this work.
\end{acknowledgements}

\bibliography{bibliography}

\providecommand{\WileyBibTextsc}{}
\let\textsc\WileyBibTextsc
\providecommand{\othercit}{}
\providecommand{\jr}[1]{#1}
\providecommand{\etal}{~et~al.}


\begin{thebibliography}{[10]}

\othercit
\bibitem{reichl_modern_1980}% book
 \textsc{L.~Reichl},
A {Modern} {Course} in {Statistical} {Physics} (Arnold, London, 1980).


\bibitem{jarzynski_nonequilibrium_1997}% article
 \textsc{C.~Jarzynski}\iffalse Nonequilibrium {Equality} for {Free} {Energy}
  {Differences}\fi,
 \jr{Phys. Rev. Lett.} \textbf{78}(14), 2690--2693 (1997).


\bibitem{crooks_entropy_1999}% article
 \textsc{G.\,E. Crooks}\iffalse Entropy production fluctuation theorem and the
  nonequilibrium work relation for free energy differences\fi,
 \jr{Phys. Rev. E} \textbf{60}(3), 2721--2726 (1999).


\bibitem{crooks_nonequilibrium_1998}% article
 \textsc{G.\,E. Crooks}\iffalse Nonequilibrium {Measurements} of {Free}
  {Energy} {Differences} for {Microscopically} {Reversible} {Markovian}
  {Systems}\fi,
 \jr{Journal of Statistical Physics} \textbf{90}(5-6), 1481--1487 (1998).


\bibitem{campisi_fluctuation_2009}% article
 \textsc{M.~Campisi},  \textsc{P.~Talkner},  and
  \textsc{P.~H{\"a}nggi}\iffalse Fluctuation {Theorem} for {Arbitrary} {Open}
  {Quantum} {Systems}\fi,
 \jr{Phys. Rev. Lett.} \textbf{102}(21), 210401 (2009).


\bibitem{talkner_tasaki_2007}% article
 \textsc{P.~Talkner} and  \textsc{P.~H{\"a}nggi}\iffalse The {Tasaki} {Crooks}
  quantum fluctuation theorem\fi,
 \jr{J. Phys. A: Math. Theor.} \textbf{40}(26), F569 (2007).


\bibitem{albash_fluctuation_2013}% article
 \textsc{T.~Albash},  \textsc{D.\,A. Lidar},  \textsc{M.~Marvian},  and
  \textsc{P.~Zanardi}\iffalse Fluctuation theorems for quantum processes\fi,
 \jr{Phys. Rev. E} \textbf{88}(3), 032146 (2013).


\bibitem{rastegin_jarzynski_2014}% article
 \textsc{A.\,E. Rastegin} and  \textsc{K.~{\.Z}yczkowski}\iffalse Jarzynski
  equality for quantum stochastic maps\fi,
 \jr{Phys. Rev. E} \textbf{89}(1), 012127 (2014).


\bibitem{collin_verification_2005}% article
 \textsc{D.~Collin},  \textsc{F.~Ritort},  \textsc{C.~Jarzynski},
  \textsc{S.\,B. Smith},  \textsc{I.~Tinoco},  and
  \textsc{C.~Bustamante}\iffalse Verification of the {Crooks} fluctuation
  theorem and recovery of {RNA} folding free energies\fi,
 \jr{Nature} \textbf{437}(7056), 231--234 (2005).


\bibitem{saira_test_2012}% article
 \textsc{O.\,P. Saira},  \textsc{Y.~Yoon},  \textsc{T.~Tanttu},
  \textsc{M.~M{\"o}tt{\"o}nen},  \textsc{D.\,V. Averin},  and  \textsc{J.\,P.
  Pekola}\iffalse Test of the {Jarzynski} and {Crooks} {Fluctuation}
  {Relations} in an {Electronic} {System}\fi,
 \jr{Phys. Rev. Lett.} \textbf{109}(18), 180601 (2012).


\bibitem{liphardt_equilibrium_2002}% article
 \textsc{J.~Liphardt},  \textsc{S.~Dumont},  \textsc{S.\,B. Smith},
  \textsc{I.~Tinoco},  and  \textsc{C.~Bustamante}\iffalse Equilibrium
  {Information} from {Nonequilibrium} {Measurements} in an {Experimental}
  {Test} of {Jarzynski}'s {Equality}\fi,
 \jr{Science} \textbf{296}(5574), 1832--1835 (2002).


\bibitem{kung_irreversibility_2012}% article
 \textsc{B.~K{\"u}ng},  \textsc{C.~R{\"o}ssler},  \textsc{M.~Beck},
  \textsc{M.~Marthaler},  \textsc{D.\,S. Golubev},  \textsc{Y.~Utsumi},
  \textsc{T.~Ihn},  and  \textsc{K.~Ensslin}\iffalse Irreversibility on the
  {Level} of {Single}-{Electron} {Tunneling}\fi,
 \jr{Phys. Rev. X} \textbf{2}(1), 011001 (2012).


\bibitem{carberry_fluctuations_2004}% article
 \textsc{D.\,M. Carberry},  \textsc{J.\,C. Reid},  \textsc{G.\,M. Wang},
  \textsc{E.\,M. Sevick},  \textsc{D.\,J. Searles},  and  \textsc{D.\,J.
  Evans}\iffalse Fluctuations and {Irreversibility}: {An} {Experimental}
  {Demonstration} of a {Second}-{Law}-{Like} {Theorem} {Using} a {Colloidal}
  {Particle} {Held} in an {Optical} {Trap}\fi,
 \jr{Phys. Rev. Lett.} \textbf{92}(14), 140601 (2004).


\bibitem{blickle_thermodynamics_2006}% article
 \textsc{V.~Blickle},  \textsc{T.~Speck},  \textsc{L.~Helden},
  \textsc{U.~Seifert},  and  \textsc{C.~Bechinger}\iffalse Thermodynamics of a
  {Colloidal} {Particle} in a {Time}-{Dependent} {Nonharmonic} {Potential}\fi,
 \jr{Phys. Rev. Lett.} \textbf{96}(7), 070603 (2006).


\bibitem{wang_experimental_2002}% article
 \textsc{G.\,M. Wang},  \textsc{E.\,M. Sevick},  \textsc{E.~Mittag},
  \textsc{D.\,J. Searles},  and  \textsc{D.\,J. Evans}\iffalse Experimental
  {Demonstration} of {Violations} of the {Second} {Law} of {Thermodynamics} for
  {Small} {Systems} and {Short} {Time} {Scales}\fi,
 \jr{Phys. Rev. Lett.} \textbf{89}(5), 050601 (2002).


\bibitem{douarche_experimental_2005}% article
 \textsc{F.~Douarche},  \textsc{S.~Ciliberto},  \textsc{A.~Petrosyan},  and
  \textsc{I.~Rabbiosi}\iffalse An experimental test of the {Jarzynski} equality
  in a mechanical experiment\fi,
 \jr{EPL} \textbf{70}(5), 593 (2005).


\bibitem{koski_distribution_2013}% article
 \textsc{J.\,V. Koski},  \textsc{T.~Sagawa},  \textsc{O.\,P. Saira},
  \textsc{Y.~Yoon},  \textsc{A.~Kutvonen},  \textsc{P.~Solinas},
  \textsc{M.~M{\"o}tt{\"o}nen},  \textsc{T.~Ala-Nissila},  and  \textsc{J.\,P.
  Pekola}\iffalse Distribution of entropy production in a single-electron
  box\fi,
 \jr{Nat Phys} \textbf{9}(10), 644--648 (2013).


\bibitem{hofmann_equilibrium_2016}% article
 \textsc{A.~Hofmann},  \textsc{V.\,F. Maisi},  \textsc{C.~R{\"o}ssler},
  \textsc{J.~Basset},  \textsc{T.~Kr{\"a}henmann},  \textsc{P.~M{\"a}rki},
  \textsc{T.~Ihn},  \textsc{K.~Ensslin},  \textsc{C.~Reichl},  and
  \textsc{W.~Wegscheider}\iffalse Equilibrium free energy measurement of a
  confined electron driven out of equilibrium\fi,
 \jr{Phys. Rev. B} \textbf{93}(3), 035425 (2016).


\bibitem{an_experimental_2015}% article
 \textsc{S.~An},  \textsc{J.\,N. Zhang},  \textsc{M.~Um},  \textsc{D.~Lv},
  \textsc{Y.~Lu},  \textsc{J.~Zhang},  \textsc{Z.\,Q. Yin},  \textsc{H.\,T.
  Quan},  and  \textsc{K.~Kim}\iffalse Experimental test of the quantum
  {Jarzynski} equality with a trapped-ion system\fi,
 \jr{Nat Phys} \textbf{11}(2), 193--199 (2015).


\bibitem{batalhao_experimental_2014}% article
 \textsc{T.\,B. Batalhão},  \textsc{A.\,M. Souza},  \textsc{L.~Mazzola},
  \textsc{R.~Auccaise},  \textsc{R.\,S. Sarthour},  \textsc{I.\,S. Oliveira},
  \textsc{J.~Goold},  \textsc{G.~De~Chiara},  \textsc{M.~Paternostro},  and
  \textsc{R.\,M. Serra}\iffalse Experimental {Reconstruction} of {Work}
  {Distribution} and {Study} of {Fluctuation} {Relations} in a {Closed}
  {Quantum} {System}\fi,
 \jr{Physical Review Letters} \textbf{113}(14) (2014).


\bibitem{jarzynski_equalities_2011}% article
 \textsc{C.~Jarzynski}\iffalse Equalities and {Inequalities}: {Irreversibility}
  and the {Second} {Law} of {Thermodynamics} at the {Nanoscale}\fi,
 \jr{Annual Review of Condensed Matter Physics} \textbf{2}(1), 329--351 (2011).


\bibitem{schleser_time-resolved_2004}% article
 \textsc{R.~Schleser},  \textsc{E.~Ruh},  \textsc{T.~Ihn},
  \textsc{K.~Ensslin},  \textsc{D.\,C. Driscoll},  and  \textsc{A.\,C.
  Gossard}\iffalse Time-resolved detection of individual electrons in a quantum
  dot\fi,
 \jr{Applied Physics Letters} \textbf{85}(11), 2005--2007 (2004).


\bibitem{vandersypen_real-time_2004}% article
 \textsc{L.\,M.\,K. Vandersypen},  \textsc{J.\,M. Elzerman},  \textsc{R.\,N.
  Schouten},  \textsc{L.\,H.\,W.\,v. Beveren},  \textsc{R.~Hanson},  and
  \textsc{L.\,P. Kouwenhoven}\iffalse Real-time detection of single-electron
  tunneling using a quantum point contact\fi,
 \jr{Applied Physics Letters} \textbf{85}(19), 4394--4396 (2004).


\bibitem{beenakker_theory_1991}% article
 \textsc{C.\,W.\,J. Beenakker}\iffalse Theory of {Coulomb}-blockade
  oscillations in the conductance of a quantum dot\fi,
 \jr{Phys. Rev. B} \textbf{44}(4), 1646--1656 (1991).


\othercit
\bibitem{ihn_semiconductor_2009}% book
 \textsc{T.~Ihn},
Semiconductor {Nanostructures}: {Quantum} states and electronic transport
  (Oxford University Press, Oxford, 2010).


\bibitem{ciorga_addition_2000}% article
 \textsc{M.~Ciorga},  \textsc{A.\,S. Sachrajda},  \textsc{P.~Hawrylak},
  \textsc{C.~Gould},  \textsc{P.~Zawadzki},  \textsc{S.~Jullian},
  \textsc{Y.~Feng},  and  \textsc{Z.~Wasilewski}\iffalse Addition spectrum of a
  lateral dot from {Coulomb} and spin-blockade spectroscopy\fi,
 \jr{Phys. Rev. B} \textbf{61}(24), R16315--R16318 (2000).


\bibitem{tarucha_shell_1996}% article
 \textsc{S.~Tarucha},  \textsc{D.\,G. Austing},  \textsc{T.~Honda},
  \textsc{R.\,J. van\,der Hage},  and  \textsc{L.\,P. Kouwenhoven}\iffalse
  Shell {Filling} and {Spin} {Effects} in a {Few} {Electron} {Quantum}
  {Dot}\fi,
 \jr{Phys. Rev. Lett.} \textbf{77}(17), 3613--3616 (1996).


\bibitem{fujisawa_bidirectional_2006}% article
 \textsc{T.~Fujisawa},  \textsc{T.~Hayashi},  \textsc{R.~Tomita},  and
  \textsc{Y.~Hirayama}\iffalse Bidirectional {Counting} of {Single}
  {Electrons}\fi,
 \jr{Science} \textbf{312}(5780), 1634--1636 (2006).


\bibitem{maclean_energy-dependent_2007}% article
 \textsc{K.~MacLean},  \textsc{S.~Amasha},  \textsc{I.\,P. Radu},
  \textsc{D.\,M. Zumb{\"u}hl},  \textsc{M.\,A. Kastner},  \textsc{M.\,P.
  Hanson},  and  \textsc{A.\,C. Gossard}\iffalse Energy-{Dependent} {Tunneling}
  in a {Quantum} {Dot}\fi,
 \jr{Phys. Rev. Lett.} \textbf{98}(3), 036802 (2007).


\bibitem{pekola_work_2012}% article
 \textsc{J.\,P. Pekola} and  \textsc{O.\,P. Saira}\iffalse Work, {Free}
  {Energy} and {Dissipation} in {Voltage} {Driven} {Single}-{Electron}
  {Transitions}\fi,
 \jr{J Low Temp Phys} \textbf{169}(1-2), 70--76 (2012).


\bibitem{averin_statistics_2011}% article
 \textsc{D.\,V. Averin} and  \textsc{J.\,P. Pekola}\iffalse Statistics of the
  dissipated energy in driven single-electron transitions\fi,
 \jr{EPL} \textbf{96}(6), 67004 (2011).


\bibitem{nyquist_thermal_1928}% article
 \textsc{H.~Nyquist}\iffalse Thermal {Agitation} of {Electric} {Charge} in
  {Conductors}\fi,
 \jr{Phys. Rev.} \textbf{32}(1), 110--113 (1928).


\bibitem{shirts_equilibrium_2003}% article
 \textsc{M.\,R. Shirts},  \textsc{E.~Bair},  \textsc{G.~Hooker},  and
  \textsc{V.\,S. Pande}\iffalse Equilibrium {Free} {Energies} from
  {Nonequilibrium} {Measurements} {Using} {Maximum}-{Likelihood} {Methods}\fi,
 \jr{Phys. Rev. Lett.} \textbf{91}(14), 140601 (2003).


\bibitem{maragakis_bayesian_2008}% article
 \textsc{P.~Maragakis},  \textsc{F.~Ritort},  \textsc{C.~Bustamante},
  \textsc{M.~Karplus},  and  \textsc{G.\,E. Crooks}\iffalse Bayesian estimates
  of free energies from nonequilibrium work data in the presence of instrument
  noise\fi,
 \jr{The Journal of Chemical Physics} \textbf{129}(2), 024102 (2008).


\end{thebibliography}
\bibliographystyle{apsrev4-1}

\end{document}